\documentclass[english,11pt,nofootinbib,showpacs,notitlepage]{revtex4-1}
\usepackage[T1]{fontenc}
\usepackage{babel}
\usepackage[utf8]{inputenc}
\usepackage{amssymb,mathrsfs,amsfonts,amsmath,euscript,textcomp,graphicx,wrapfig,color,txfonts,lmodern,multirow,dsfont}
\usepackage[section]{placeins}
\usepackage[usenames,dvipsnames,svgnames,table]{xcolor}
\definecolor{darkgreen}{rgb}{0,.7,0}
\usepackage[colorlinks,linkcolor=blue,citecolor=darkgreen,pagebackref=true]{hyperref}
\def\dac{\displaystyle\frac}

\def\[{\left[}
\def\]{\right]}
\def\({\left(}
\def\){\right)}

\newcommand{\eq}[1]{\begin{equation}#1\end{equation}}

\newcommand{\oneN}{=\overline{1,N}}

\newcommand{\const}{\mathop{\rm const}\nolimits}

\newcommand{\e}{\mathop{\rm e}\nolimits}

\newcommand{\bgp}[1]{\bigl(#1\bigr)}

\begin{document}
\baselineskip7mm

\title{Exact exponential solutions in Einstein-Gauss-Bonnet flat anisotropic cosmology}

\author{Dmitry Chirkov}
\affiliation{Sternberg Astronomical Institute, Moscow State University, Moscow 119991 Russia}
\affiliation{Faculty of Physics, Moscow State University, Moscow 119991 Russia}
\author{Sergey Pavluchenko}
\affiliation{Instituto de Ciencias F\'isicas y Matem\'aticas, Universidad Austral de Chile, Valdivia, Chile}
\author{Alexey Toporensky}
\affiliation{Sternberg Astronomical Institute, Moscow State University, Moscow 119991 Russia}

\begin{abstract}
In this paper we are looking for the exponential solutions (i.e. the solutions with the scale factors change exponentially over time) in the
Einstein-Gauss-Bonnet gravity. We argue that we found all possible non-constant-volume
solutions (up to permutations) in lower dimensions
((4+1) and (5+1)) and developed a scheme which allows one to find necessary conditions in arbitrary dimension.
\end{abstract}


\maketitle

\section{Introduction}
Due to the developments in the string and related theories the interest to the multidimensional gravity was increasing rapidly in the last decades. Indeed,
the gravitational counterpart of these theories allows one to perform some kind of independent viability test on them. Thinking of possible implications of
gravitational theories one can notice that the observational data on the compact objects (including what is believed to be ``black holes'') is still too
unprecise to make differences between different gravitational theories -- not to mention, the existence of the black holes is still have not been proven.
On the other hand, the gravitational waves are also beyond our reach -- let us hope, just for now; these two facts make cosmology the best (if not the only)
candidate to test the viability of the modified gravity. Mainly because of this, since 1980s the studies of the modified multidimensional cosmology were
quite intense (see e.g.
\cite{add_1, add_2, add_3, add_4, Deruelle1, add_6, add_7, add_8, Deruelle2, add_10, add13, add_11, add_12, mpla09, Pavluchenko, grg10, KPT, prd10, PT, CGP}).

When doing multidimensional cosmology, it is naturally to do it anisotropic -- indeed, if we want to link originally multidimensional universe to our observed
(3+1)-dimensional Universe, one have to think of a way to reduce the number of dimensions -- a way to compactify -- and this is clear indication of anisotropy.
And there are successful attempts to describe ``natural'' compactification -- due to dynamical~\cite{add13} or topological~\cite{CGP} effects. Yet, studies of
the generic anisotropic solutions is also important since with wider analysis we get better understanding of the dynamics.

Studies of anisotropic cosmological solutions in multidimensional
Gauss-Bonnet (GB) gravity reveals several important differences between
GB and General Relativity (GR). More than 20 years ago it was shown that
Kasner solution of GR is replaced by similar power-law solution with different relations
between power indices~\cite{Deruelle1, Deruelle2}.

What is more interesting, it appears
that the form of solution similar to the Kasner solution in GR is not unique for
a flat anisotropic background in GB. Another possible solution with two {\it arbitrary}
power indices (with others equal to zero) have been obtained in the above cited paper.
No counterpart of this solution exists in GR. In~\cite{Pavluchenko} such solution have
been generalized for higher Lovelock terms. It is interesting, however, that though
such solutions containing unconstrained values of appropriate number of power indices
are exact solutions in Lovelock gravity when only the highest Lovelock term is taken
into account, they disappear when subdominant terms are added~\cite{PT}.

Recently another type of vacuum solution in GB gravity have been found. In contrast to
any analogue of the Kasner solution, the scale factors in the solution in question change
exponentially~\cite{Ivashchuk}. Again, no analogue of this solution exists in GR.
In~\cite{KPT} this solution have been generalized to non-vacuum case, though detailed
analysis of its existence have not been provided. In the present paper we study the conditions
for the exponential solution to exist in non-vacuum case. In addition, we provide full analysis of exponential solutions in $(4+1)$ and $(5+1)$ dimensions in models with a cosmological
constant, generalizing particular results of \cite{KPT}.

The structure of the manuscript is as follows: in the second section we present the equations of motion for the system under consideration, in third --
consider the case in low ((4+1) and (5+1)) number of dimensions, in the following section we generalize some of the solutions obtained on the higher number
of dimensions and finally in the Conclusions sum up the obtained results.

\section{The set-up.}

We start from the Einstein-Gauss-Bonnet action in $(N+1)$-dimensional spacetime\footnote{Throughout the paper we use the system of units in which $G=c=1$, $G$ is the $(N+1)$-dimensional gravitational constant, $c$ is the speed of light. Greek indices run from 0 to N, Latin indices from 1 to N.}:
\eq{S=\frac{1}{16\pi}\int d^{N+1}x\sqrt{|\det(g)|}\bgp{\mathcal{L}_E+\alpha\mathcal{L}_{GB}+\mathcal{L}_m},\quad\mathcal{L}_E=R,\quad
\mathcal{L}_{GB}=R_{\alpha\beta\gamma\delta}R^{\alpha\beta\gamma\delta}-4R_{\alpha\beta}R^{\alpha\beta}+R^2,\label{action}}
where $R,R_{\alpha\beta},R_{\alpha\beta\gamma\delta}$ are the $(N+1)$-dimensional scalar curvature, Ricci tensor and Riemann tensor respectively; $\mathcal{L}_m$ is the Lagrangian of a matter; $\det(g)$ is the determinant of a metric tensor $g$. We introduce the metric components as
\eq{g_{00}=-1,\quad g_{kk}=\e^{2H_kt},\quad g_{ij}=0,\;\;i\ne j.}
In the following we interest in the case of $H_i\equiv\const$. In this work we deal with a perfect fluid with the equation of state $p=\omega\rho$ as a matter source. Let $\varkappa=8\pi\rho$; the equations of motions now can be written as follows:
\eq{2\sum\limits_{i\ne j}H_i^2+2\sum\limits_{\{i>k\}\ne j}H_i H_k+8\alpha\sum\limits_{i\ne j}H_i^2\sum\limits_{\{k>l\}\ne\{i, j\}}H_k H_l+24\alpha\sum\limits_{\{i>k>l>m\}\ne j}H_i H_k H_l H_m=-\omega\varkappa,\quad j\oneN;\label{eq.of.motion}}
\eq{2\sum\limits_{i>j}H_i H_j+24\alpha\sum\limits_{i>j>k>l}H_i H_j H_k H_l=\varkappa.\label{constraint}}
Since we have $H_i\equiv\const$, it follows from Eqs.~(\ref{eq.of.motion})--(\ref{constraint}) that $\rho\equiv\const$, so that the continuity equation
\eq{\dot\rho+(\rho+p)\sum_i H_i=0}
reduces to
\eq{(\rho+p)\sum_i H_i=0,}
which allows several different cases: a) $\rho\equiv0$ (vacuum case), b) $\rho+p=0$ ($\Lambda$-term case), c) $\sum_i H_i=0$ (constant volume case) and
their combinations: d) $\sum_i H_i=0$ vacuum and e) $\sum_i H_i=0$ with $\Lambda$-term. In present paper we do not consider constant volume solutions (CVS),
leaving their description for a separate paper. So we left with only options -- either vacuum ($\rho = 0$) or $\Lambda$-term ($\rho + p = 0$) and further we
consider only these two cases.

\section{Low-dimensional cases.\label{vacuum.and.Lambda.term}}

Subtracting $i$-th dynamical equation from $j$-th one we obtain:
\eq{\(H_j - H_i\)\(\frac{1}{4\alpha}+\sum\limits_{\{k>l\}\ne\{ i, j\}}H_k H_l\)\sum_k H_k=0\label{after.subtraction}}
It follows from~(\ref{after.subtraction}) that
\eq{H_{i}=H_{j}\quad\vee\quad \sum\limits_{\{k>l\} \ne \{ i, j\}} H_k H_l=-\frac{1}{4\alpha}\quad\vee\quad\sum_k H_k=0}
These are necessary conditions for a given set $H_1,\ldots,H_N$ to be a solution of Eqs.~(\ref{eq.of.motion})--(\ref{constraint}).
The case $\sum_k H_k=0$ is CVS and will be considered in a separate paper; in this section we deal with the following possibilities only:
\eq{H_{i}=H_{j}\quad\vee\quad \sum\limits_{\{k>l\} \ne \{ i, j\}} H_k H_l=-\frac{1}{4\alpha}\label{disj}}
We call the left equality as type I condition, the right equality as type II condition.

\subsection{(4+1)-dimensional spacetime.}
In this case we have 6 pairs of the type I and type II conditions. The table~\ref{table.cond} lists all such pairs. We assign numbers from 1 to 6 to each type I condition and letters from $\mathcal{A}$ to $\mathcal{F}$ to each type II condition. There are the following combinations of type I and type II conditions:
\begin{enumerate}
  \item 0 type I conditions, 6 type II conditions $\Longrightarrow\;H_1=H_2=H_3=H_4\equiv H=\frac{1}{\sqrt{-4\alpha}}$;
  \item 1 (any) type I conditions, 5 type II conditions $\Longrightarrow\;H_1=H_2=H_3=H_4\equiv H=\frac{1}{\sqrt{-4\alpha}}$;
  \item 2 type I conditions, 4 type II conditions:
  \begin{enumerate}
    \item type I conditions has no identical parameters, for example, (see table~\ref{table.cond}) $1-2-\mathcal{B}-\mathcal{D}-\mathcal{E}-\mathcal{F}\;\Longrightarrow$ $H_1=H_2\equiv H,\,H_3=H_4\equiv h,\;Hh=-\frac{1}{4\alpha}$;
    \item both of type I conditions include one the same parameter, for example, $1-3-\mathcal{C}-\mathcal{D}-\mathcal{E}-\mathcal{F}\;\Longrightarrow$ $H_1=H_2=H_3=H_4\equiv H=\frac{1}{\sqrt{-4\alpha}}$;
  \end{enumerate}
  \item 3 type I conditions, 3 type II conditions:
  \begin{enumerate}
    \item a chain of conditions does not include one of the parameters, for example, $1-3-5-\mathcal{C}-\mathcal{D}-\mathcal{F}$ ($H_4$ is absent) $\Longrightarrow\;H_1=H_2=H_3\equiv H=\frac{1}{\sqrt{-\alpha}},\,H_4\equiv h\in\mathds{R}$,
    \item a chain of conditions includes all the parameters $\Longrightarrow\;H_1=H_2=H_3=H_4\equiv H=\frac{1}{\sqrt{-4\alpha}}$;
  \end{enumerate}
  \item 4 (or 5) type I conditions, 2 (or 1) type II conditions $\Longrightarrow\;H_1=H_2=H_3=H_4\equiv H=\frac{1}{\sqrt{-4\alpha}}$;
  \item 6 type I conditions, 0 type II conditions $\Longrightarrow\;H_1=H_2=H_3=H_4\equiv H\in\mathds{R}$.
\end{enumerate}

\begin{table}[t]
\begin{center}
  \caption{The necessary conditions.}
  \label{table.cond}
  \begin{tabular}{|c|c|c|c||c|c|c|c|}
  \hline
    & type I &   & type II & & type I &   & type II \\
  \hline
  1 & $H_1=H_2$ & $\mathcal{A}$ & $H_3H_4=-\frac{1}{4\alpha}$ & 4 & $H_1=H_4$ & $\mathcal{D}$ & $H_2H_3=-\frac{1}{4\alpha}$ \\
  \hline
  2 & $H_3=H_4$ & $\mathcal{B}$ & $H_1H_2=-\frac{1}{4\alpha}$ & 5 & $H_2=H_3$ & $\mathcal{E}$ & $H_1H_4=-\frac{1}{4\alpha}$ \\
  \hline
  3 & $H_1=H_3$ & $\mathcal{C}$ & $H_2H_4=-\frac{1}{4\alpha}$ & 6 & $H_2=H_4$ & $\mathcal{F}$ & $H_1H_3=-\frac{1}{4\alpha}$ \\
  \hline
\end{tabular}
\end{center}
\end{table}
Summarizing aforesaid we see that there are three cases: i) $H_1=H_2=H_3=H_4\equiv H\in\mathds{R}$ (clearly, $H=\frac{1}{\sqrt{-4\alpha}}$ is the subcase of that case); ii) $H_1=H_2=H_3\equiv H=\frac{1}{\sqrt{-4\alpha}},\,H_4\equiv h\in\mathds{R}$; iii) $H_1=H_2\equiv H,\,$ $H_3=H_4\equiv h,\;Hh=-\frac{1}{4\alpha}$. In $(4+1)$-dimensional spacetime Eqs.~(\ref{eq.of.motion})--(\ref{constraint}) have the form:
\eq{2\sum_{i\ne j}^4H_i^2+2\sum_{\{i>k\}\ne j}^4H_i H_k+8\alpha\sum_{i\ne j}^4H_i^2\sum_{\{k>l\}\ne\{i, j\}}^4H_k H_l=-\omega \varkappa,\quad j=\overline{1,4}\label{eq.of.motion.4+1}}
\eq{2\sum_{i>j}^4H_i H_j+24\alpha H_1 H_2 H_3 H_4=\varkappa\label{constraint.4+1}}
Now we consider solutions of the Eqs.~(\ref{eq.of.motion.4+1})--(\ref{constraint.4+1}) for each of the cases i)--iii).

\textbf{i}) After substitution $H_1=H_2=H_3=H_4\equiv H$ into Eqs.~(\ref{eq.of.motion.4+1})--(\ref{constraint.4+1}) one can see that they reduce to
\eq{12H^2+24\alpha H^4+\omega\varkappa=0,\quad 12H^2+24\alpha H^4-\varkappa=0\label{4+1.isotropic}}
one can see that the system~(\ref{4+1.isotropic}) could be resolved only if $\varkappa=0$ (vacuum case) or $\omega=-1$ ($\Lambda$-case); both of these cases could be reduced to the following solution:
\eq{H^2=\frac{1}{4\alpha}\left(-1\pm\sqrt{1+\frac{2\varkappa\alpha}{3}}\right),\quad
    \left[\begin{array}{c}
            \varkappa\geqslant-\frac{3}{2\alpha},\;\alpha>0 \\
            \varkappa\leqslant-\frac{3}{2\alpha},\;\alpha<0
          \end{array}
    \right.}
For vacuum case ($\varkappa=0$) we obtain $H=0$ or $H^2=-\frac{1}{2\alpha},\,\alpha<0$. It is interesting to note that the cases \textbf{1}, \textbf{2}, \textbf{3b}, \textbf{4b} and \textbf{5} are realized as a subcase of the case iii) for $\Lambda=-\frac{3}{16\pi\alpha},\;\alpha<0$ (see below).

\textbf{ii}) Substitution $H_1=H_2=H_3\equiv H=\frac{1}{\sqrt{-4\alpha}},\,H_4\equiv h\in\mathds{R}$ into Eqs.~(\ref{eq.of.motion.4+1})--(\ref{constraint.4+1}) gives a system which could be resolved only under $\omega=-1$, which gives us $-\frac{3}{2\alpha}=\varkappa$.

\textbf{iii}) Substitution $H_{1}=H_{2}\equiv H,\,H_{3}=H_{4}\equiv h$ with the condition $Hh=-\frac{1}{4\alpha}$ into Eqs.~(\ref{eq.of.motion.4+1})--(\ref{constraint.4+1}) leads to
a system that could be resolved only under $\omega=-1$ which could be further brought to
\eq{H^2+h^2=\frac{\varkappa}{2}+\frac{1}{4\alpha}\label{H^2+H h+h^2}}
It follows from~(\ref{H^2+H h+h^2}) immediately that $\varkappa>-\frac{1}{2\alpha}$. Substituting $h=-\frac{1}{4\alpha H}$ into~(\ref{H^2+H h+h^2}) we obtain
\eq{H^4-\left(\frac{\varkappa}{2}+\frac{1}{4\alpha}\right)H^2+\frac{1}{16\alpha^2}=0\label{H^2+H h+h^2.after.subs}}
Solutions of Eq.~(\ref{H^2+H h+h^2.after.subs}) are
\eq{H^2=\frac{1}{4}\left(\varkappa+\frac{1}{2\alpha}\pm\sqrt{\left(\varkappa+\frac{1}{2\alpha}\right)^2-\frac{1}{\alpha^2}}\right),
\quad\varkappa\geqslant\frac{1}{|\alpha|}-\frac{1}{2\alpha}\;\vee\;\varkappa\leqslant-\frac{1}{|\alpha|}-\frac{1}{2\alpha}\label{2+2}}

One can easily see that in vacuum case ($\varkappa=0$) the radical in (\ref{2+2}) is negative so such solution do not exist in that case. We summarize the results in the tables~\ref{table.solutions.alpha.pos},\ref{table.solutions.alpha.neg}.

\begin{table}[h]
\begin{center}
\caption{$\Lambda$-term solutions, $\alpha>0$.}
\label{table.solutions.alpha.pos}
  \begin{tabular}{|c|c|c|}
    \hline
    & $0<\Lambda\leqslant\frac{1}{16\pi\alpha}$ & $\Lambda>\frac{1}{16\pi\alpha}$  \\
    \hline
    $(H,H,h,h)$ & No & $H^2=\frac{1}{4}\left(8\pi\Lambda+\frac{1}{2\alpha}\pm\sqrt{\left(8\pi\Lambda+\frac{1}{2\alpha}\right)^2-\frac{1}{\alpha^2}}\right), h=-\frac{1}{4\alpha H}$ \\
    \hline
    $(H,H,H,h)$ & \multicolumn{2}{c|}{No} \\
    \hline
    $(H,H,H,H)$ & \multicolumn{2}{c|}{$H^2=\frac{1}{4\alpha}\left(-1+\sqrt{1+\frac{16\pi\alpha\Lambda}{3}}\right)$}  \\
    \hline
  \end{tabular}
\end{center}
\end{table}
\begin{table}[h]
\begin{center}
\caption{$\Lambda$-term solutions, $\alpha<0$.}
\label{table.solutions.alpha.neg}
  \begin{tabular}{|c|c|c|c|c|}
    \hline
    & $\Lambda<-\frac{3}{16\pi\alpha}$ & $\Lambda=-\frac{3}{16\pi\alpha}$ & $\Lambda>-\frac{3}{16\pi\alpha}$ \\
    \hline
    $(H,H,h,h)$ & No & $H^2=h^2=-\frac{1}{4\alpha}$ & $\begin{array}{c}
            H^2=\frac{1}{4}\left(8\pi\Lambda+\frac{1}{2\alpha}\pm\sqrt{\left(8\pi\Lambda+\frac{1}{2\alpha}\right)^2-\frac{1}{\alpha^2}}\right),\\
            h=-\frac{1}{4\alpha H}
          \end{array}$ \\
    \hline
    $(H,H,H,h)$ & No & $\begin{array}{c} H^2=-\frac{1}{4\alpha}, \\ h\in\mathds{R} \end{array}$ & No \\
    \hline
    $(H,H,H,H)$ & $\begin{array}{c}H^2=\frac{1}{4\alpha}\left(-1\pm\sqrt{1+\frac{16\pi\alpha\Lambda}{3}}\right) \\
                           \mbox{solutions with positive square root} \\
                           \mbox{exist for}\; \Lambda<0\; \mbox{only}
           \end{array}$ & $H^2=-\frac{1}{4\alpha}$ & No \\
    \hline
  \end{tabular}
\end{center}
\end{table}

The vacuum case admits only isotropic solution: $H^2=-\frac{1}{2\alpha},\;\alpha<0$. Let us note that solution with one the Hubble parameter different from the other three ($H_1=H_2=H_3=H,\,H_4=h$) exists for the single value of $\Lambda$: $\Lambda=-\frac{3}{16\pi\alpha},\;\alpha<0$.
Also one can note that ($H,\,H,\,h,\,h$) solution tends smoothly to the  isotropic solution for $\Lambda=-\frac{3}{16\pi\alpha}$ and negative $\alpha$.

\subsection{(5+1)-dimensional spacetime.}
In this case we have 10 pairs of the type I and type II conditions~(\ref{disj}). Note that, as distinct from the (4+1)-dimensional case, type II conditions are the sums of three pairwise products of Hubble parameters but it does not affect follow-up reasoning: as in  the (4+1)-dimensional case one can check that there are the following necessary conditions for a given set of the Hubble parameters $H_1,\ldots,H_5$ to be a solution of the dynamical equations: i) $H_1=H_2=H_3=H_4=H_5\equiv H$; ii) $H_1=H_2=H_3=H_4\equiv H,\,H_5\equiv h$; iii) $H_1=H_2=-H_3=-H_4\equiv H,\,H_5\equiv h$; iv) $H_1=H_2=H_3\equiv H,\,$ $H_4=H_5\equiv h$.
Similarly to (4+1)-dimensional case, i) could be resolved only if $\varkappa=0$ or $\omega=-1$; ii)--iv) have solution with $\omega=-1$ and the particular case
of $\sum H=0$ (once they fulfill the conditions for the solutions of this branch, it is enough for us for now); additionally, iv) has vacuum ($\varkappa=0$) solution.

\begin{table}[h]
\begin{center}
\caption{$\Lambda$-term solutions.}
\label{table.solutions.5+1}
  \begin{tabular}{|c|c|c|}
    \hline
    & $\alpha>0$ & $\alpha<0$ \\
    \hline
    $(H,H,H,H,H)$ & $H^2=\frac{1}{12\alpha}\left(-1+\sqrt{1+\frac{48\pi\alpha\Lambda}{5}}\right),\;\Lambda>0$ &
    $\begin{array}{c}
       H^2=\frac{1}{12\alpha}\left(-1-\sqrt{1+\frac{48\pi\alpha\Lambda}{5}}\right),\;\Lambda<-\frac{5}{48\pi\alpha} \\
       H^2=\frac{1}{12\alpha}\left(-1+\sqrt{1+\frac{48\pi\alpha\Lambda}{5}}\right),\;\Lambda<0
     \end{array}
    $ \\
    \hline
    $(H,H,H,H,h)$ & No & $H^2=-\frac{1}{12\alpha},\;h\in\mathds{R},\;\Lambda=-\frac{5}{48\pi\alpha}$ \\
    \hline
    $(H,H,-H,-H,h)$ & $H^2=\frac{1}{4\alpha},\;h\in\mathds{R},\;\Lambda=\frac{1}{16\pi\alpha}$ & No \\
    \hline
  \end{tabular}
\end{center}
\end{table}

Solutions for $\Lambda$ cases of i),ii),iii) are presented in the table~\ref{table.solutions.5+1}. The last case iv) is more complicated; it leads to the following equations:
\eq{192\,{H}^{6}{\alpha}^{3}-112\,{H}^{4}{\alpha}^{2}+ \left( 256\,\Lambda\,\pi \,{\alpha}+4\right){H}^{2}\alpha-1=0,\quad h=-{\frac {4\,{H}^{2}\alpha+1}{8H\alpha}}\label{3+2}}
It is easy to check that when $\alpha>0$ Eqs.~(\ref{3+2}) has at least one solution for any $\Lambda$; when $\alpha<0$ Eqs.~(\ref{3+2}) has at least one solution iff $\Lambda>-\frac{5}{48\pi\alpha}$.
Additionally, if we require $\Lambda > 0$ (which seems more physically motivated) we can impose an additional constraint on the solution
of the Eq. (\ref{3+2}). In that case we require $(192\xi^3 - 112\xi^2 + 4\xi - 1) < 0$ where $\xi = \alpha H^2$. This equation has only one real root
$\xi_0 = (4\sqrt[3]{10} + \sqrt[3]{100} + 7)/36 \approx 0.56276$, so that the constraint is $\alpha H^2 < \xi_0$.

The vacuum case admits:
\begin{itemize}
  \item isotropic solution: $H^2=-\frac{1}{6\alpha},\;\alpha<0$;
  \item solution with (3+2) spatial splitting: $H_1=H_2=H_3=H,\;H_4=H_5=\xi H$ and
\eq{H^2=\frac{f(\xi)}{4\alpha}\Biggl|_{\xi=\frac{\sqrt[3]{10}}{3}-\frac{\sqrt[3]{100}}{6}-\frac{2}{3}\approx -0.722}\;,\quad f(\xi)=-\frac{\xi^2+6\xi+3}{3\xi(3\xi+2)}\label{32vac}}
\end{itemize}
As well as in the (4+1)-dimensional case solution with one the Hubble parameter different from the others ($H_1=H_2=H_3=H_4=H,\,H_5=h$) exists for the single value of $\Lambda$: $\Lambda=-\frac{5}{48\pi\alpha},\;\alpha<0$.

\section{Higher dimensions}
Equation (\ref{after.subtraction}) is valid for any number of dimensions and it is necessary (but not sufficient) condition for solution to exist. Previously
we found all solutions with non-constant volume in (4+1) and (5+1) dimensions, but formally we can extend an analysis to higher dimensions as well. ``Formally''
because there are doubts about using Gauss-Bonnet in higher dimensions -- from one hand, in (6+1) and higher in the spite of Lovelock theory cubic correction
comes into play and equations of motion become more complicated. However, on the other hand it is Gauss-Bonnet but not Lovelock gravity which is motivated by the string
theory, so some motivation for considering higher dimensional EGB gravity still remains.

There are two cases for the Hubble parameters distribution -- all of them could be distinct or there could be multiple values. In the first case neither of $H_i-H_j$ terms could be zero and only the case with a special sum of pairwise products remains. In addition to these sums, if we sum all possible pairwise products, we get one more condition:

\begin{equation}
\begin{array}{l}
\sum\limits_{i>j} H_i H_j = - \dac{D(D-1)}{4\alpha (D-2) (D-3)}.
\end{array} \label{sum_diff}
\end{equation}

The case with possible repetitions in Hubble parameters is a bit more tricky: difference equations for multiple values are fulfilled since $(H_i-H_i)=0$, so what remain are the difference
equations for distinct Hubble parameters, whose number is lower than the number of the dimensions. Thus the difference equations for the case with multiple values for Hubble parameters
are equivalent to the case with subset of these Hubble parameters made of all distinct Hubble parameters. For instance, the case of $D$ dimensions with $D$ distinct Hubble parameters
($H_1,\dots,\,H_D$) and the case of $(D+1)$ with two Hubble parameters are equal to each other (say, $H_1,\,H_1,\dots,\,H_D$ -- so that the number of distinct Hubble parameters is $D$)
are equivalent to each other in the terms of the difference equation. And now let us stress an attention again on the fact that these are necessary conditions -- because the sufficient
conditions are different for the example considered above.

Generally, with increase in the number of dimensions the analysis becomes more and more complicated for each particular case, making finding
solutions a difficult task. Yet, some of the solutions found could be generalized. One of them is the isotropic solution; substituting
$H_i \equiv H$ into (\ref{eq.of.motion})--(\ref{constraint}) we obtain two independent equations

\begin{equation}
\begin{array}{l}
D (D-1) H^2 + \alpha D (D-1)(D-2)(D-3) H^4 = \varkappa,\quad D (D-1) H^2 + \alpha D (D-1)(D-2)(D-3) H^4 = - \omega\varkappa;
\end{array} \label{iso_high}
\end{equation}

\noindent similar to the lower-dimensional cases one can see that (\ref{iso_high}) could be resolved only in either vacuum ($\varkappa = 0$) or in
$\Lambda$ ($\omega = -1$) cases. The former of them admits either trivial or

\begin{equation}
\begin{array}{l}
H^2 = -\dac{1}{\alpha (D-2)(D-3)}
\end{array} \label{iso_high_H}
\end{equation}

\noindent solution; one can easily verify that (\ref{iso_high_H}) is also valid in lower-dimensional cases. The $\Lambda$-solution of (\ref{iso_high})
could be split into two subcases -- with positive or negative $\alpha$. For $\alpha > 0$ the solution for $H$ and constraint on $\Lambda$ are

\begin{equation}
\begin{array}{l}
H^2 = \dac{\sqrt{1+32\alpha\pi\Lambda \dac{(D-2)(D-3)}{D(D-1)}} -1}{2\alpha (D-2)(D-3)};\quad \Lambda > 0.
\end{array} \label{iso_high_L1}
\end{equation}

\noindent For $\alpha < 0$ we have up to two branches

\begin{equation}
\begin{array}{l}
H^2 = \dac{\pm\sqrt{1+32\alpha\pi\Lambda \dac{(D-2)(D-3)}{D(D-1)}} -1}{2\alpha (D-2)(D-3)};\quad \Lambda < - \dac{D(D-1)}{32\pi\alpha(D-2)(D-3)};
\end{array} \label{iso_high_L2}
\end{equation}

\noindent and ``+'' branch exists only if $\Lambda < 0$. One can easily check that expressions in (\ref{iso_high_L1}) and (\ref{iso_high_L2}) coincide
with those in lower dimensions (tables \ref{table.solutions.alpha.pos}, \ref{table.solutions.alpha.neg}, and \ref{table.solutions.5+1}).

The second solution which could be generalized is the solution with $(D-1)+1$ spatial splitting -- i.e.
with Hubble parameters in the form $(H,\,H,\,\dots,\, H,\,h)$. This solution is

\begin{equation}
\begin{array}{l}
H^2 = - \dac{1}{2\alpha (D-2)(D-3)},~\Lambda = - \dac{D(D-1)}{32\alpha\pi (D-2)(D-3)}.
\end{array} \label{split_high_L}
\end{equation}

\noindent Once again one can make sure that lower-dimensional values from tables \ref{table.solutions.alpha.pos}, \ref{table.solutions.alpha.neg},
and \ref{table.solutions.5+1} meet general expressions in (\ref{split_high_L}) and the solution follows the same rules: it exist only
for a single value of $\Lambda$, only for $\alpha < 0$ and $h\in\mathds{R}$.

Let us note that the general higher-dimensional case do not, of course, limits with above-mentioned solutions -- but only these two could be easily
generalized for higher dimensional case.

\section{Conclusions}

We argue that we obtained all possible non-constant-volume exponential solution in the Einstein-Gauss-Bonnet gravity in (4+1) and (5+1) dimensions.
Some of these solutions were generalized on the arbitrary number of dimensions.

The starting point is the Eq. (\ref{after.subtraction}) which determine the necessary conditions for the solutions to exist. There we can clearly separate
constant volume case (last
condition) and consider it separately -- we are going to do so in incoming paper. The first two conditions become (\ref{disj}) and in lower number of
dimensions ((4+1) and (5+1)) they could be solved explicitly.
It appears that the sector with negative $\alpha$ is more abundant with the solutions -- say, vacuum isotropic or $\Lambda$-solutions with $((D-1)+1)$ spatial
splitting could be obtained only if $\alpha < 0$ in any number of dimensions; for other solutions generally the case with $\alpha < 0$ allows wide variety of
branches. This could explain why in previous numerical studies~\cite{mpla09, grg10, KPT, prd10} so little number of ``good'' solutions were found --
most of them correspond to $\alpha < 0$ case which is believed to possess serious problems on a quantum level.

In the higher number of dimensions (\ref{disj}) become more complicated so we decided to leave complete investigation for further analysis. Still, some of the
solutions retrieved for (4+1) and (5+1) could be generalized for any number of dimensions, and that was done as well.

\textit{Acknowledgments.-- }
This work was supported by RFBR grant No. 14-02-00894. S.A.P. was supported by FONDECYT via grant No. 3130599.

\end{document}